\documentclass[prb,11pt,onecolumn]{revtex4}

\usepackage{graphicx,amssymb,amsmath}

\begin{document}

\author{Haining Wang}
\email{hw343@cornell.edu}
\author{Changjian Zhang}
\author{Farhan Rana}
\affiliation{School of Electrical and Computer Engineering, Cornell University, Ithaca, NY, USA}

\title{Ultrafast Dynamics of Defect-Assisted Electron-Hole Recombination in Monolayer MoS$_{2}$}

\begin{abstract}
In this letter, we present non-degenerate ultrafast optical pump-probe studies of the carrier recombination dynamics in MoS$_{2}$ monolayers. By tuning the probe to wavelengths much longer than the exciton line, we make the probe transmission sensitive to the total population of photoexcited electrons and holes. Our measurement reveals two distinct time scales over which the photoexcited electrons and holes recombine; a fast time scale that lasts $\sim$2 ps and a slow time scale that lasts longer than $\sim$100 ps. The temperature and the pump fluence dependence of the observed carrier dynamics are consistent with defect-assisted recombination as being the dominant mechanism for electron-hole recombination in which the electrons and holes are captured by defects via Auger processes. Strong Coulomb interactions in two dimensional atomic materials, together with strong electron and hole correlations in two dimensional metal dichalcogenides, make Auger processes particularly effective for carrier capture by defects. We present a model for carrier recombination dynamics that quantitatively explains all features of our data for different temperatures and pump fluences. The theoretical estimates for the rate constants for Auger carrier capture are in good agreement with the experimentally determined values. Our results underscore the important role played by Auger processes in two dimensional atomic materials. 
\end{abstract}

\maketitle

\begin{figure}[tbh]
  \begin{center}
   \includegraphics[width=0.7\textwidth]{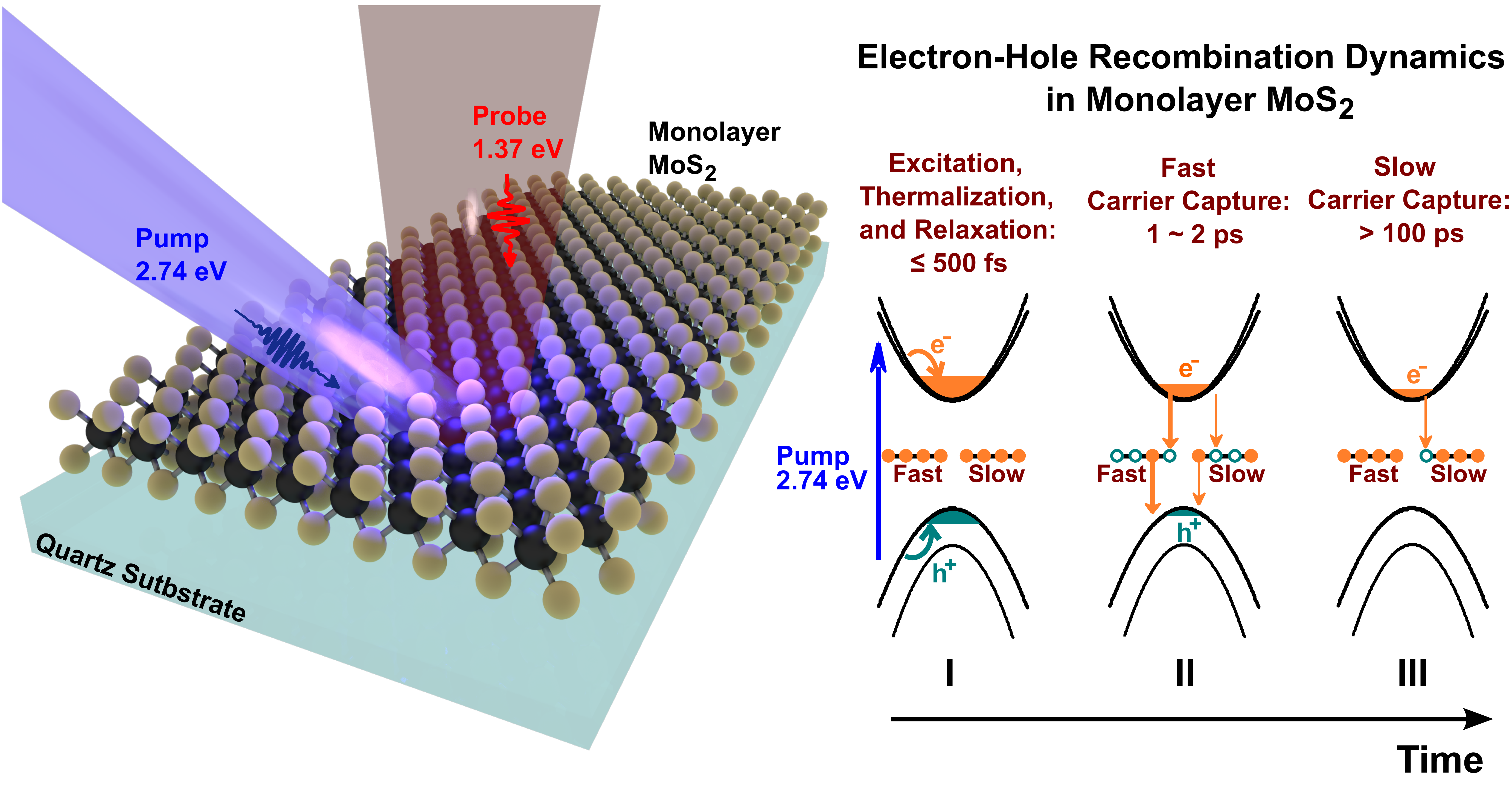}
    \label{pump}
  \end{center}
\end{figure}

Two-dimensional (2D) transition metal dichalcogenides (TMDs) have emerged as interesting materials both from the perspective of basic science as well as applications~\cite{Podzorov04,Splendiani10,Mak10,VanDerZande13,Wang12,Mak12,KFM13}. Applications of these materials in electronics and optoelectronics have been extensively explored in recent years~\cite{Wang12,Mak12,Yoon11,WangH12,Lopez13,Zhu14,Kis11,Ross14,Iwasa12,Das13,Hua12,Steiner13,Baugher14}. The bandgaps of most TMDs are in the visible to near-IR wavelength range, making these materials suitable for light emitters, photodetectors, and solar cells~\cite{Wang12,Lopez13,Ross14,Hua12,Steiner13,Baugher14}. In addition, optical control of valley polarization in TMDs has opened opportunities for devices based on the valley degree of freedom~\cite{Mak12}. The lifetimes of electrons and holes are critical to all the proposed and demonstrated TMD optoelectronic devices. Despite the recent progress, carrier lifetimes and non-radiative electron-hole recombination mechanisms in TMDs remain poorly understood. Developing a better understanding of the non-radiative electron-hole recombination mechanisms in TMDs is especially important because the reported quantum efficiencies in both TMD light emitters and detectors are extremely poor; in the .0001-.01 range~\cite{Lopez13,Hua12,Ross14,Steiner13,Baugher14}. Similar quantum efficiencies for TMDs have been observed in photoluminescence experiments~\cite{Splendiani10,Mak10}. In contrast, the best reported internal and external quantum efficiencies observed in photoluminescence in III-V semiconductors exceed 0.9 and 0.7, respectively~\cite{Schnit93}. Therefore, most of the electrons and holes injected either electrically or optically in TMDs recombine non-radiatively. The mechanisms by which electrons and holes recombine non-radiatively, and the associated time scales, remain to be clarified.

In this letter, we report results on the ultrafast dynamics of photoexcited carriers from non-degenerate optical pump-probe experiments performed on monolayer MoS$_{2}$. In contrast to the previously reported optical pump-probe studies of monolayer MoS$_{2}$~\cite{Huang13,WangR12,Choi13}, in which the probe wavelength was tuned close to the exciton line ($\sim$650 nm), the probe wavelength in our experiments is chosen to be much longer than the exciton line such that the probe transmission is affected predominantly by intraband absorption from the electrons and holes created by the pump pulse and not by resonant or near-resonant interband excitonic nonlinearities which are more difficult to interpret quantitatively~\cite{Chemla85}. The probe transmission in our experiments is used to observe the total photoexcited carrier populations, including both free carriers and bound carriers (excitons), and their dynamics. Our results are consistent with defect-assisted recombination as being the dominant mechanism for electron-hole recombination in which the the electrons and holes are captured by defects via Auger processes. The temperature and the pump fluence dependence of the observed carrier dynamics, together with the small photoluminescence quantum efficiencies, rule out most other recombination mechanisms as playing the dominant role. In most bulk semiconductors, Auger processes for carrier capture by defects are believed to be important at high carrier densities and single and multi-phonon processes for carrier capture dominate at low carrier densities~\cite{Lang77,Lax60,Ridley13,Landsberg92,Landsberg80}. However, electron-electron and electron-hole interactions are particularly strong in two-dimensional TMDs. For example, the exciton binding energies in TMDs are almost two orders of magnitude larger than in most III-V semiconductors~\cite{Changjian14,Berk13}. The strong Coulomb interactions in TMDs, including correlations in the positions of electrons and holes arising from the attractive interaction, result in large carrier capture rates via Auger scattering (theoretical details are provided in the supplementary information). Our results show that the decay transients of the photoexcited carrier density are not simple exponentials and exhibit different time scales. The measured dynamics and time scales can be explained quantitatively for all temperatures and for all pump fluences by assuming electron and hole capture by defects with different capture rates via Auger scattering.                                

\begin{figure}[tbh]
  \centering
  \includegraphics[width=.7\textwidth]{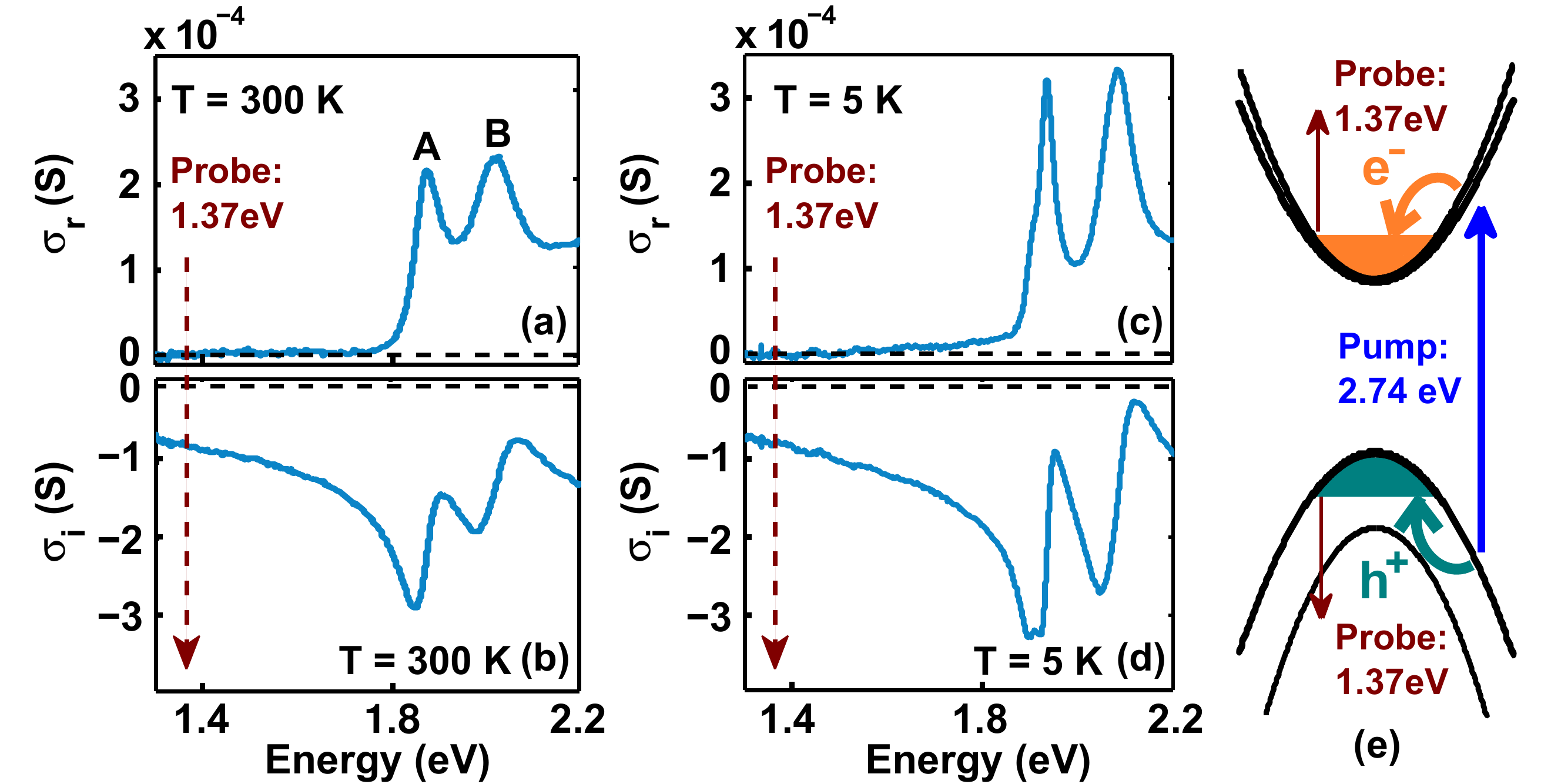}
  \caption
  {(a,b,c,d) The measured real and imaginary parts of the optical conductivity of monolayer MoS$_2$ sample are plotted for 5 K and 300 K. The $A$ and $B$ exciton resonances as well as the probe photon energy are also indicated. (e) Schematic of the optical pump-probe scheme is shown. The probe differential transmission is affected predominantly by changes in the real part of the  optical conductivity due to intraband absorption (intra-conduction band and intra-valence band absorption).}
  \label{fig:trans}
\end{figure}

\noindent
\textbf{Sample Preparation and Experimental Technique:} Our monolayer samples were mechanically exfoliated from bulk MoS$_2$ (obtained from SPI Supplies and 2D Semiconductors) and transferred onto quartz substrates. Sample thickness was confirmed by both Raman and transmission/reflection spectroscopy and monolayers were identified~\cite{Ryu10}. The samples were found to be n-doped. The electron density was estimated from Raman measurements to be in the 2-3$\times10^{12}$ 1/cm$^{2}$ range~\cite{Changjian14,sood12}. Electrical measurements on similar samples on oxide-coated doped silicon substrates (with electrostatic gating) yielded intrinsic electron densities in the same range~\cite{Changjian14}. The samples were placed in a helium-flow cryostat. In the optical pump-probe experiments, $\sim$80 fs pulses at 905 nm wavelength (1.37 eV photon energy) from a Ti-Sapphire laser were frequency doubled to 452 nm (2.74 eV) by a beta-BaB$_{2}$O$_{4}$ crystal. The 2.74 eV pump pulses were used to excite electrons from the valence band into the conduction band in the sample, and the differential transmission ($\Delta T/T$) of the time-delayed 1.37 eV pulses were used to probe the sample after photoexcitation. A 20X objective was used to focus the pump and the probe beams onto the sample. From direct pump absorption measurements, 1 $\mu$J/cm$^{2}$ pump fluence is estimated to generate an electron (and hole) density of $\sim$2.5$\times$10$^{11}$ 1/cm$^{2}$. The measurement time resolution was $\sim$400 fs, and was limited by the dispersion of the optics in the setup. The measured $\Delta T/T$ can be expressed as\cite{Dawlaty08,George08},
\begin{equation} 
\frac{\Delta T}{T} \approx -2\eta_{o}\frac{\Delta\sigma_{r}}{1+n_{s}} - 2 \eta_{o}^{2} \frac{(\sigma_{r}\Delta\sigma_{r}+\sigma_{i}\Delta\sigma_{i})}{(1+n_{s})^{2}} \label{eq:1}
\end{equation}  
Where, $\sigma_{r}$ ($\sigma_{i}$) is the real (imaginary) part of the sample optical conductivity, $n_{s}\approx 1.45$ is the refractive index of the substrate, and $\eta_{o}$ is the free-space impedance. The optical conductivities were measured by broadband transmission and reflection spectroscopies, and are shown in Figures \ref{fig:trans}(a)-(d), at 5 and 300 K, respectively. When the probe energy is either near an exciton resonance or near a band-edge, the sign and the magnitude of the changes in the optical conductivities after photoexcitation, due to excitonic optical nonlinearities and band-filling effects, can have a complicated dependence on the probe energy. This makes quantitative interpretation of pump-probe data a difficult task~\cite{Chemla85}. We therefore chose the probe energy to be much smaller than the exciton resonance. No detectable optical absorption or photoluminescence was observed in the samples at the probe energy (1.37 eV) indicating that the sample had no optically active midgap defect states at this energy. Changes in the imaginary part of the optical conductivity after photoexcitation due to excitonic optical non-linearities and band filling effects are expected to be positive at the probe energy~\cite{Chemla85}, thereby making $\Delta T/T$ positive, which is contrary to our experimental observations discussed below. In addition, since $\eta_{o} \sigma_{r}$ and $\eta_{o} \sigma_{i}$ are both $\ll 1$ at the probe energy (see Figure \ref{fig:trans}), the second term on the right hand side in (\ref{eq:1}) is expected to be much smaller than the first term. The differential transmission of the probe is expected to be affected predominantly by changes in the real part of the optical conductivity as given by the first term on the right hand side in (\ref{eq:1}) due to intra-conduction band and intra-valence band absorption~\cite{scsm67}. Therefore, $\Delta T/T$ is expected to be negative.

The exciton binding energies in most TMDs are in the few hundred meV range~\cite{Changjian14,Berk13}. Such large binding energies imply that photoexcited electrons and holes could form excitons relatively quickly. The goal of our experiments is not to study the exciton formation dynamics. When the probe photon energy is much smaller than the exciton binding energies, the excitons respond like a charge neutral insulating gas and, unlike free carriers, do not contribute to the probe intraband absorption~\cite{Kaindl03,Kaindl09}. When the probe photon energy is much larger than the exciton binding energies and also much smaller than the optical bandgap, as is the case in our experiments, excitons contribute to the intraband absorption in approximately the same way as the free carriers~\cite{Kira06,Kaindl09,Olszakier89,Sadeghi04}. This is shown explicitly in the supplementary information using exciton conductivity sum rules. The change in the real part of the optical intraband conductivity of the sample at the probe frequency can therefore be written approximately in the Drude form as (see supplementary information),
\begin{equation}
\Delta \sigma_{r}(\omega) \approx \left( \frac{\Delta n}{m_{e}} + \frac{\Delta p}{m_{h}} \right) \frac{e^{2}\tau}{1 + \omega^{2} \tau^{2}} \label{eq:drude}
\end{equation}
where $\omega$ is the probe frequency, $\tau$ is the damping rate, $m_{e}$ and $m_{h}$ are the electron and hole effective masses (assumed to be equal to 0.5m$_{o}$~\cite{Changjian14}), and $\Delta n$ and $\Delta p$ are the photo-induced changes in the total electron and hole densities including both free and bound (excitons) carriers. Therefore, the probe transmission in our experiments is sensitive to the total carrier population, and enables studies of the carrier recombination dynamics and mechanisms, which is the focus of our work.

\begin{figure}[tbh]
  \centering
  \includegraphics[width=.5\textwidth]{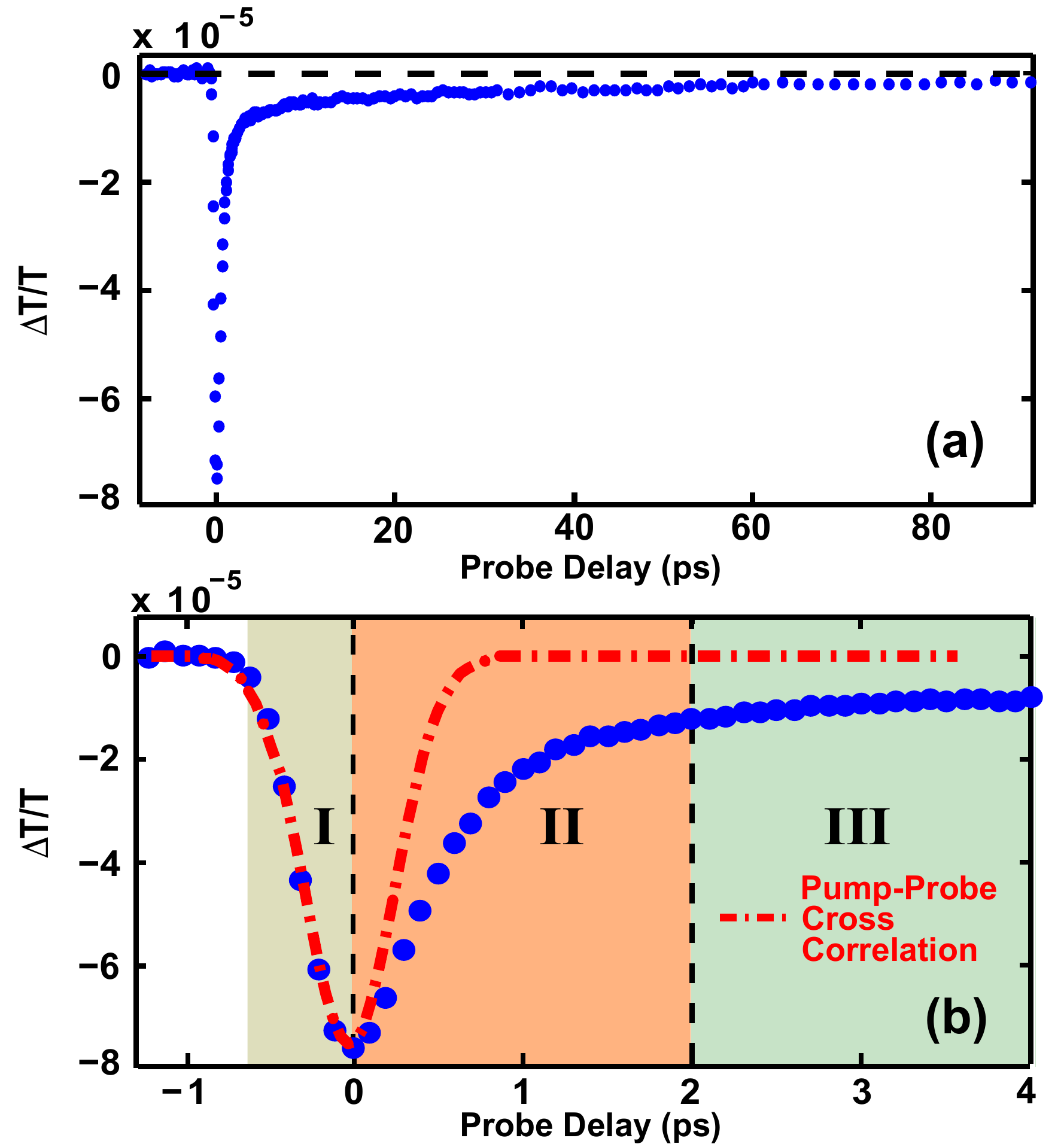}
  \caption
  {(a) The measured differential transmission $\Delta T/T$ of the probe pulse is plotted as a function of the probe delay with respect to the pump pulse. The pump fluence is $\sim$16 $\mu$J/cm$^{2}$ and T=300 K. (b) A zoomed-in plot of the data in (a) shows three different temporal regions: (I) $\Delta T/T$ reaches its negative maximum within $\sim$500 fs. (II) A fast recovery of the negative $\Delta T/T$ then occurs within $\sim$2 ps. (III) Finally, a slow recovery of the negative $\Delta T/T$ lasts for more than a hundred picoseconds.} 
  \label{fig:power}
\end{figure}

\begin{figure}[tbh]
  \centering
  \includegraphics[width=.5\textwidth]{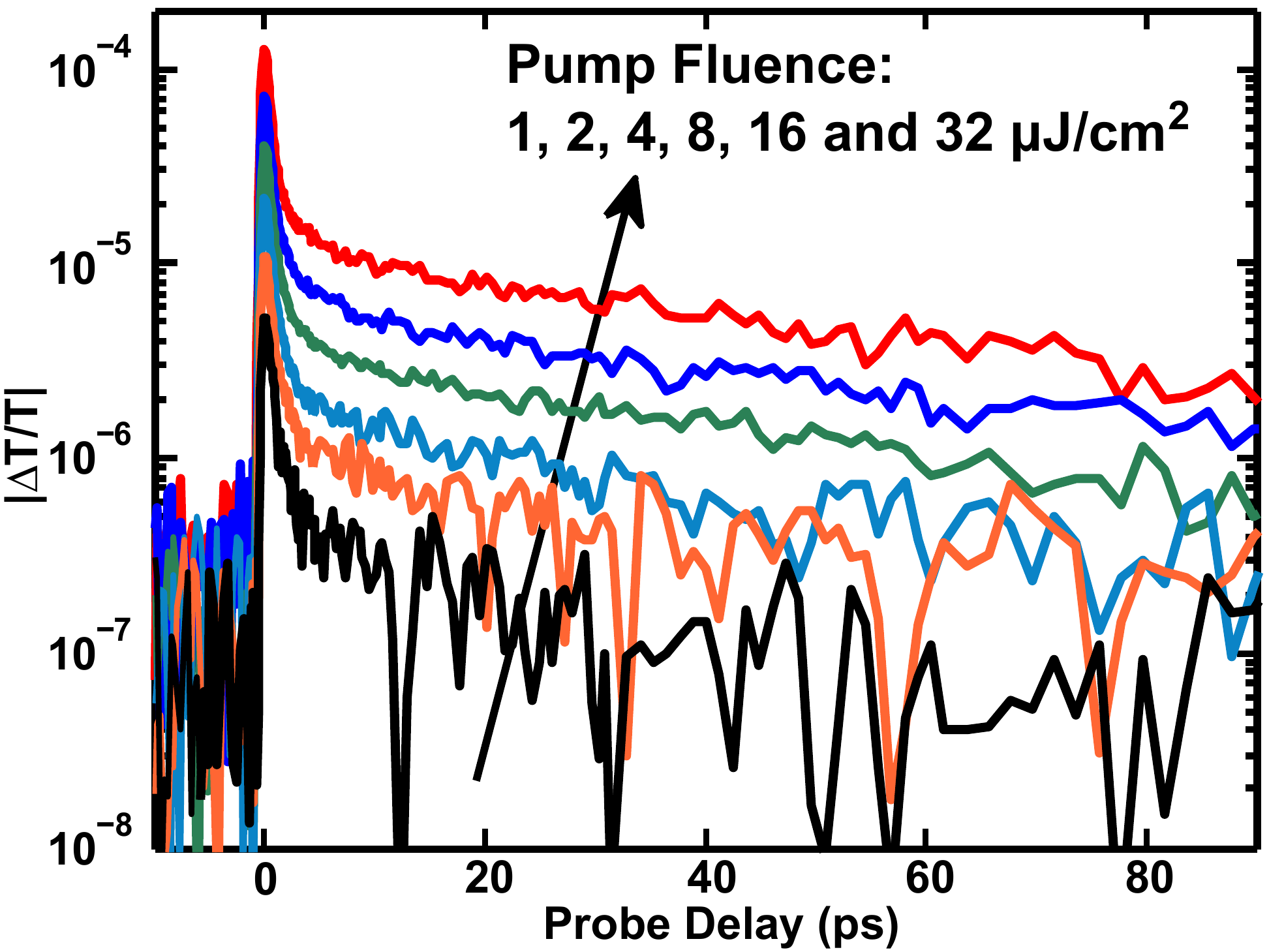}
  \caption
  {(a) The measured $|\Delta T/T|$ of the probe pulse is plotted as a function of the probe delay for different pump fluences (1, 2, 4, 8, 16, and 32 $\mu$J/cm$^{2}$) at T=300 K. The slow transient in region (III) appears to be nearly exponential with a time constant of 60-70 ps.}     
  \label{fig:log_data}
\end{figure}

\noindent
\textbf{Experimental Results and Discussion:} A differential transmission transient measured at room temperature is shown in Figure \ref{fig:power} for 16 $\mu$J/cm$^{2}$ pump fluence (T = 300 K). $\Delta T/T$ is observed to be negative, as expected from the intraband absorption of the probe photons by the photoexcited electron and hole populations. Three different time scales (or temporal regions) are observed and marked in Figure \ref{fig:power}(b): (I) Upon photoexcitation by the pump pulse, $\Delta T/T$ reaches its maximum negative value within $\sim$500 fs. (II) A fast recovery of the negative $\Delta T/T$ then occurs within $\sim$2 ps. (III) Finally, a slow recovery of the negative $\Delta T/T$ lasts for more than hundred picoseconds. Figure \ref{fig:log_data} shows $\Delta T/T$ for different pump fluences (1, 2, 4, 8, 16, and 32 $\mu$J/cm$^{2}$) at T=300 K. The slow transient in region (III) appears to be nearly exponential with a time constant of 60-70 ps. The data presented here was reproducible in different samples exfoliated from both natural and synthetic bulk crystals obtained from different vendors (SPI Supplies and 2D Semiconductors) with less than 10\% variation in the observed time scales across samples. The samples were found to be permanently damaged by pump fluence values exceeding $\sim$50 $\mu$J/cm$^{2}$. Once damaged in this way, the measured transients changed dramatically, non-repeatably, and exhibited much longer time scales (see supplementary information).      

\begin{figure}[tbh]
  \centering
  \includegraphics[width=.7\textwidth]{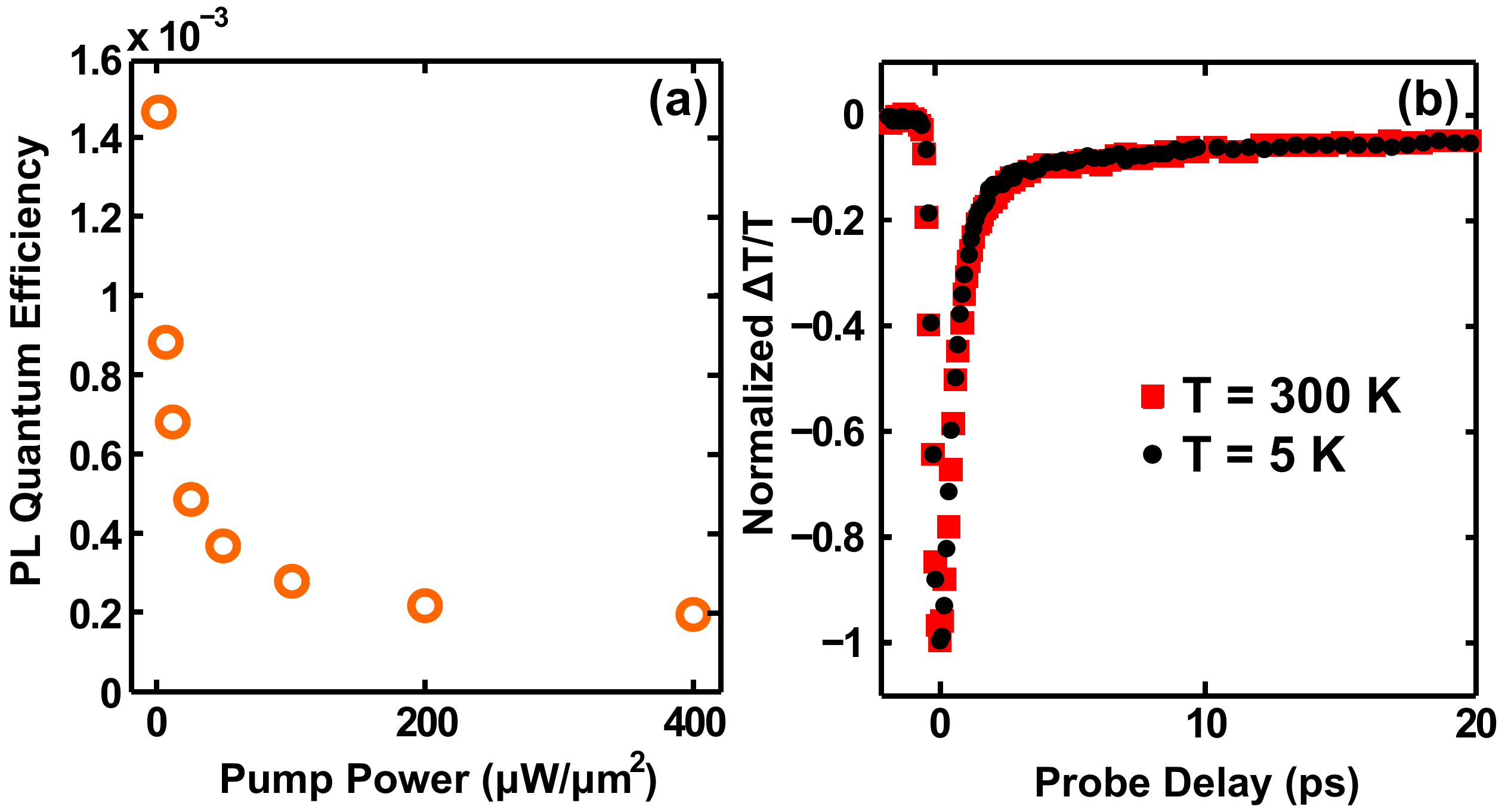}
  \caption
  {(a) Radiative quantum efficiency of a suspended monolayer MoS$_{2}$ sample, obtained by integrating the photoluminescence (PL) spectra, is plotted as a function of the optical pump power (pump wavelength was 532 nm). T=300 K. (b)  The measured $\Delta T/T$ (normalized) of the probe pulse is plotted as a function of the probe delay for two different temperatures (5 K and 300 K). The data shows no significant temperature dependence of the time scales associated with the transient. The pump fluence was $\sim$32 $\mu$J/cm$^{2}$ in both cases.}
  \label{fig:quantum}
\end{figure}
\noindent

We now discuss the processes that contribute to the observed transients. Possible thermal diffusion of hot photoexcited carriers out of the pump or probe focus spots was ruled out as a contributing factor to the observed transients by changing the focus spot size in measurements~\cite{WangR12}. Although carrier generation by the pump pulse and subsequent hot carrier intraband relaxation can contribute to the observed transient in region (I), our measurement is limited by the temporal resolution of the setup as indicated by the good fit of the transient in region (I) with the pump-probe cross-correlation curve. The asymmetry of the observed transient in (I) and (II) shows that two-photon absorption (TPA) between the pump and the probe pulses does not contribute in any significant way to the measured differential transmission of the probe pulse. The recovery of the negative $\Delta T/T$ occurs over very different time scales in regions (II) and (III). These two different time scales have been observed in previous ultrafast studies~\cite{Huang13,Schuller11,Lagarde14}. The fast initial decay in region (II) cannot be attributed to intraband thermalization or hot carrier intraband relaxation since similar fast initial decay was also observed in ultrafast PL measurements by Lagarde et~al.~\cite{Lagarde14} (limited by the $\sim$4 ps time resolution in their experiments) and PL is likely to increase as the carriers thermalize and cool. Also, the intraband carrier relaxation times for electrons were measured by Tanaka et~al.~\cite{Tanaka03} in bulk MoS$_{2}$ using two-photon photoemission spectroscopy and times shorter than $\sim$50 fs were obtained for electrons with energies a few tenths of an eV from the conduction band edge~\cite{Tanaka03}. It is therefore reasonable to assume that the photoexcited electrons and holes thermalize and lose most of their energy (via optical phonon emission) in the first few hundred femtoseconds after photoexcitation in a manner similar to what happens, for example, in graphene~\cite{Wang10}. 

The small absolute values, as well as the pump intensity dependence, of the radiative quantum efficiencies in our samples provide a clue to interpret the transient in regions (II) and (III). Figure \ref{fig:quantum}(a) shows the measured radiative quantum efficiency of a suspended monolayer MoS$_{2}$ sample, obtained by integrating the PL spectra, plotted as a function of the optical pump intensity (pump in this case was a 532 nm wavelength continuous-wave laser and T = 300 K). The quantum efficiency was estimated based on the amount of actual light collected by a 100X objective~\cite{Mak10}. The small values of the quantum efficiencies, in agreement with the previously reported values~\cite{Mak10}, show that most of the photoexcited carriers recombine non-radiatively. The decrease of the quantum efficiency with the pump intensity indicates that the steady state non-radiative recombination rates increase faster with the photoexcited electron and hole densities compared to the radiative recombination rates. In addition, radiative lifetimes of excitons and trions were recently reported by us~\cite{Wang14} and were found to be generally much longer compared to the picosecond scale dynamics observed in region (II). Based on these observations and considerations we attribute the measured transient in regions (II) and (III) to the non-radiative capture and recombination of photoexcited carriers. 

\begin{figure}[tbh]
  \centering
  \includegraphics[width=.7\textwidth]{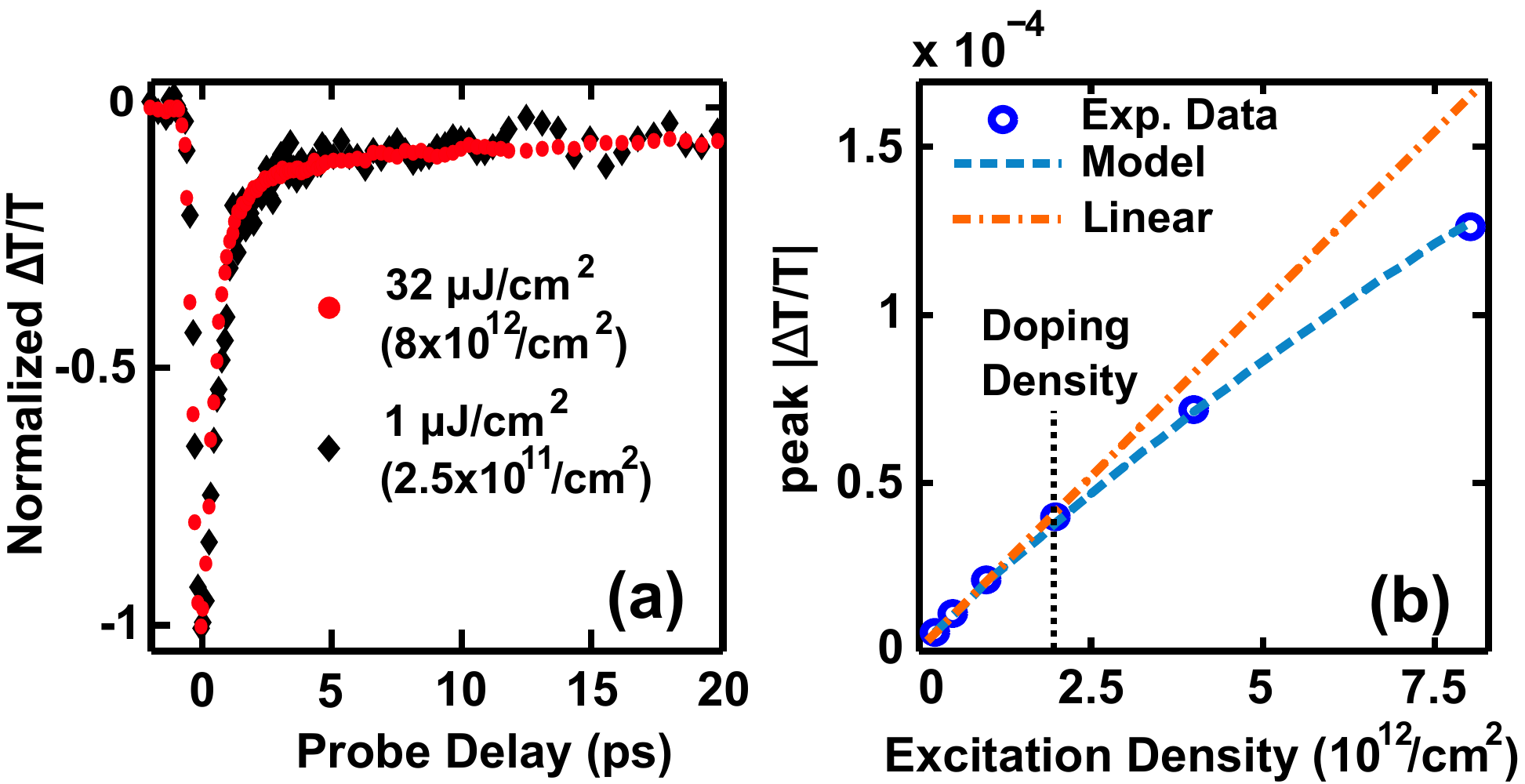}
  \caption
  {(a) $\Delta T/T$ (normalized) is plotted for two different pump fluences, 1 and 32  $\mu$J/cm$^{2}$, at T=300 K. Whereas the slow time scale in region (III) is completely independent of the pump fluence, the fast time scale in region (II) becomes marginally faster (by 10-15\%) at highest pump fluence compared to at the lowest pump fluence. (b) The peak value of $|\Delta T/T|$ is plotted as a function of the photoexcited carrier density (estimated from the pump fluence) showing a mildly sub-linear dependence. The fit obtained from the Auger carrier capture model is also shown. T=300 K.}     
  \label{fig:power_scaling}
\end{figure}

In order to determine the non-radiative capture and recombination mechanisms we look at the temperature and pump fluence dependence of the time scales observed in the transients. Interestingly, the time scales in $\Delta T/T$ exhibited no observable temperature dependence over the entire temperature range 5-300 K. In Figure \ref{fig:quantum}(b), normalized $\Delta T/T$ is plotted as a function of the probe delay for two different temperatures (5 K and 300 K) and shows no significant temperature dependence in the time scales (the pump fluence was fixed at $\sim$32 $\mu$J/cm$^{2}$ in both cases). In Figure \ref{fig:power_scaling}(a), the normalized transients for two extreme pump fluence values, 1 and 32 $\mu$J/cm$^{2}$, are plotted (T=300 K). The data shows that the time scales in the transient are largely independent of the pump fluence in the entire range of the pump fluence values used in our experiments. A more careful examination reveals that while the time scale of the slow transient in region (III) is indeed independent of the pump fluence, the time scale of the fast transient in region (II) becomes marginally faster (by 10-15\%) at highest pump fluence compared to at the lowest pump fluence. Figure \ref{fig:power_scaling}(b) shows the peak value of $|\Delta T/T|$ plotted as a function of the photoexcited carrier density (estimated from the pump fluence). $|\Delta T/T|$ shows a mildly sub-linear behavior with the pump fluence. Below we show that defect-assisted recombination via Auger carrier capture explains all features of our data.      

\begin{figure}[tbh]
  \centering
  \includegraphics[width=.5\textwidth]{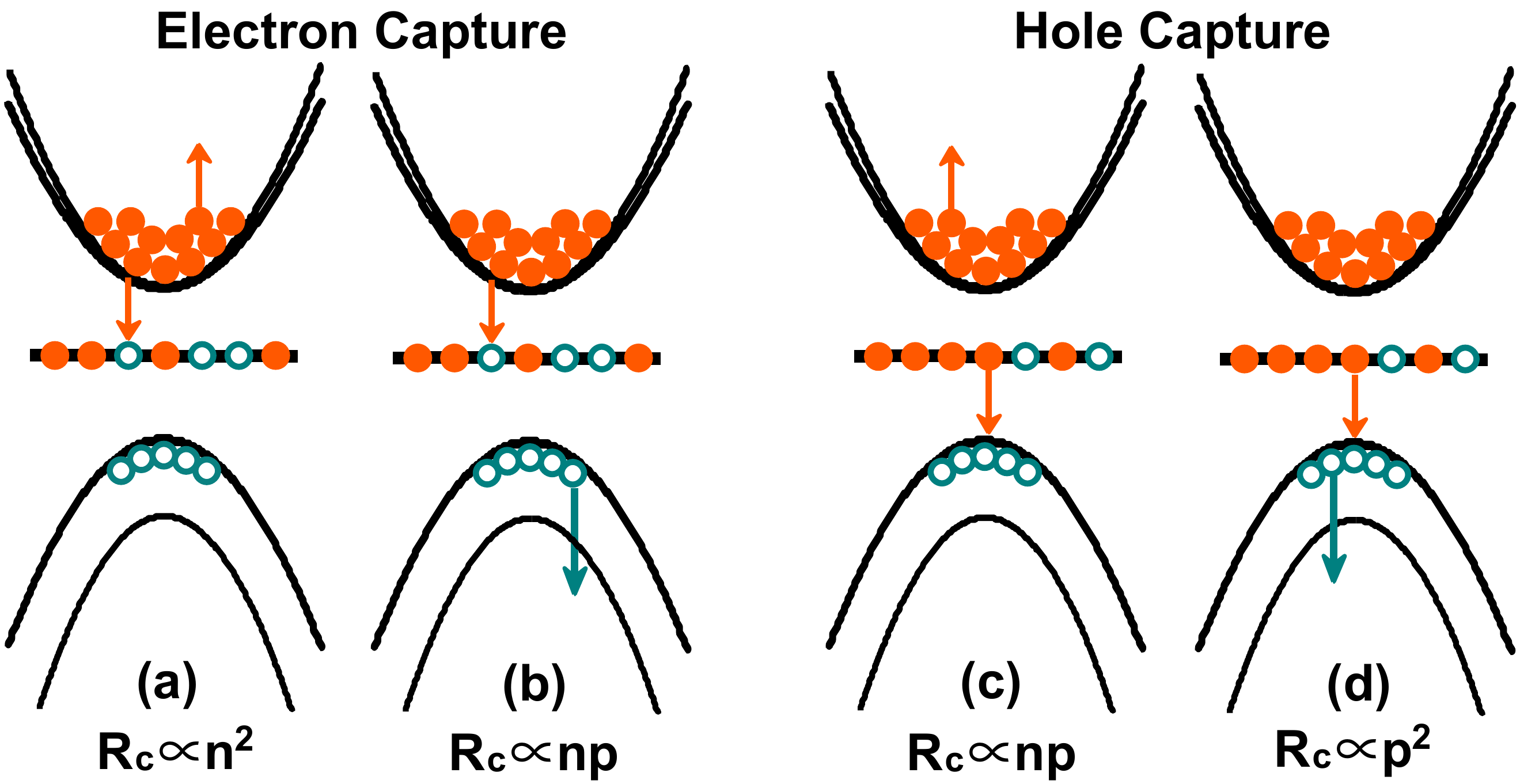}
  \caption{Four basic Auger processes for the capture of electrons ((a) and (b)) and holes ((c) and (d)) at defects are depicted. The approximate dependence of the capture rates on the electron and hole densities is also indicated for each process~\cite{Jaros80,Landsberg92,Landsberg80}. In n-doped samples, processes (a) and (c) are expected to dominate for electron and hole capture, respectively, for low to moderate photoexcited carrier densities.}  
  \label{fig:auger}
\end{figure}

\noindent
\textbf{Auger Carrier Capture Model:} Recombination via direct band-to-band Auger scattering is generally slow for large bandgap materials~\cite{Ridley13} and it also cannot explain the sharp transition in the time scales observed in the transient between regions (II) and (III). Electron and hole capture by defects in defect-assisted non-radiative recombination occurs mainly by two mechanisms: (a) phonon-assisted processes, and (b) Auger processes. Phonon-assisted processes can be single-phonon processes or multi-phonon processes, including phonon-cascade processes~\cite{Lang77,Lax60,Ridley13}. The carrier capture rates in all phonon-assisted processes depend strongly on the lattice temperature~\cite{Lang77,Lax60,Ridley13}. For example, the rates of activated multi-phonon capture processes depend exponentially on the lattice temperature~\cite{Lang77,Ridley13}. In contrast, the rate of carrier capture by defects via Auger processes is largely temperature independent and consistent with our observations~\cite{Jaros80,Landsberg92,Landsberg80,Haug93}(see also supplementary information). Figure \ref{fig:auger} shows the four basic Auger processes for the capture of electrons ((a) and (b)) and holes ((c) and (d)) at defects and the approximate dependence of the capture rates on the electron and hole densities~\cite{Jaros80,Landsberg92,Landsberg80}. The corresponding emission processes are the just the inverse of the capture processes shown. The carrier density dependence of the capture rates shown in Figure \ref{fig:auger} holds approximately for both bound (excitons) and free carriers (see supplementary information). In our n-doped samples, processs (a) and (c) are expected to dominate for electron and hole capture, respectively, at low to moderate pump fluence values. 

\begin{figure}[tbh]
  \includegraphics[width=.6\textwidth]{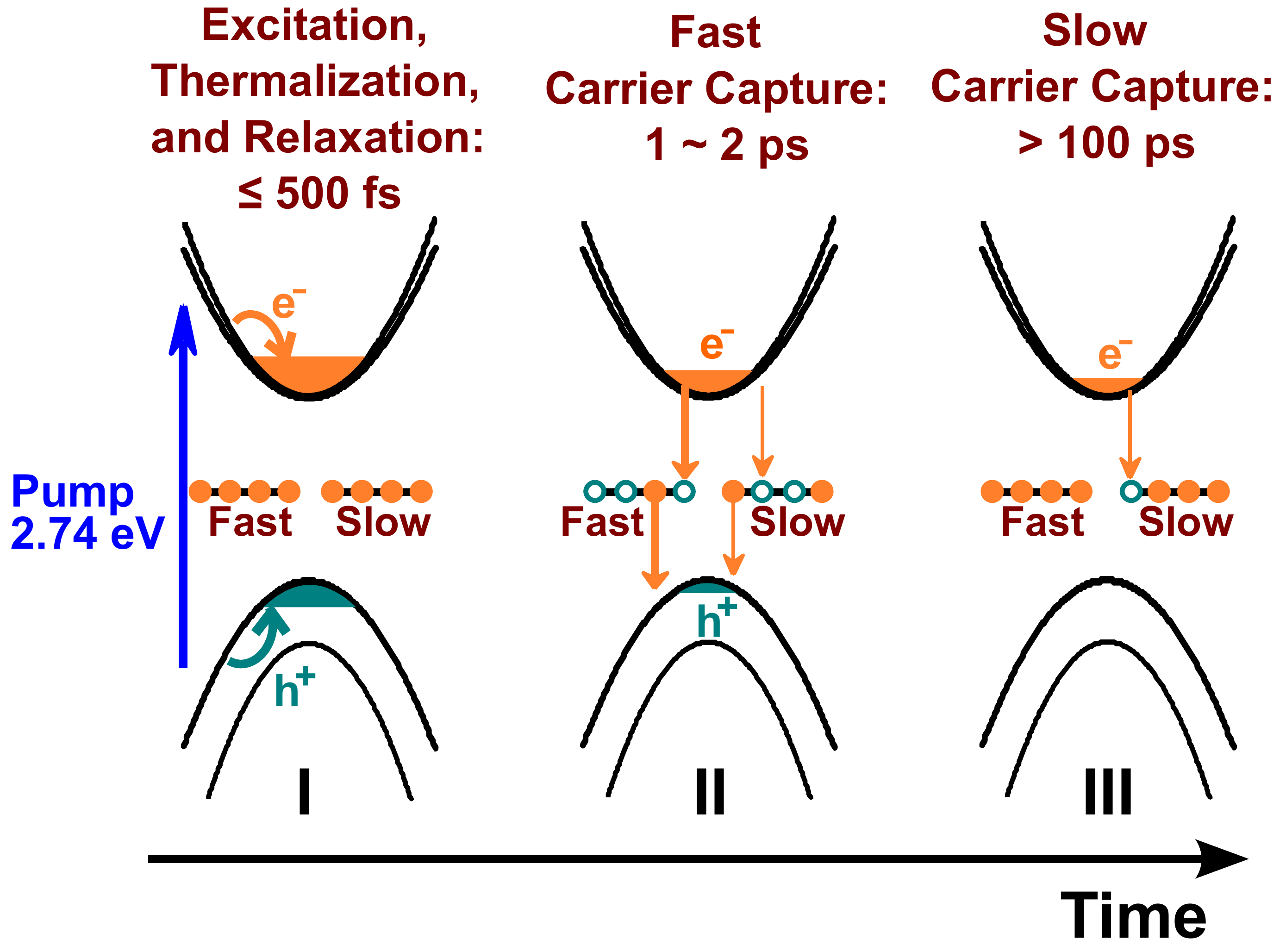}
  \caption
{Illustration of the ultrafast carrier dynamics in MoS$_{2}$ in the three temporal regions: (I) After photoexcitation, the carriers thermalize and cool and form a correlated electron-hole plasma. (II)  Most of the holes, followed by the electrons, are captured by the fast defects within 1-2 ps. A small fraction of the photogenerated holes is also captured by the slow defects. (III) After all the photoexcited holes have been captured and the electrons have filled the fast traps completely, the remaining photoexcited electrons are captured by the slow defects on a 60-70 ps time scale and the slow transient lasts for more than 100 ps.}
  \label{fig:cartoon}
\end{figure}

A simple rate equation model based on Auger carrier capture by defects can be developed that explains all features of our data. We note that the sudden transition from the fast time scale in region (II) to the slow time scale in region (III) in the measured transients cannot be explained by saturation of the defects states alone since, as already discussed, the measured time scales in Figure \ref{fig:power_scaling} do not exhibit strong pump fluence dependence. Most semiconductors contain defect levels with very different carrier capture rates~\cite{HaynesI55, HaynesII55, Blakemore58,Choo70,Wertheim58}. Monolayer MoS$_{2}$ is known to have several different kinds of point defects, such as sulfur and molybdenum vacancies and interstitials, in addition to grain boundaries and dislocations~\cite{Sofo04,Komsa12,Seifert13,Zhou13,Kim14,Guinea14,Robertson13,Hao13,VanDerZande13}. We assume two different deep midgap defect levels in our samples; a fast defect level and a slow defect level, labeled by subscripts $f$ and $s$, respectively. In our n-doped samples, the defect levels are assumed to be fully occupied in thermal equilibrium. Keeping only the most important Auger capture processes in n-doped materials ((a) and (c) in  Figure \ref{fig:auger}), and ignoring electron and hole emission from the deep defects, we obtain the following rate equations for the electron and hole densities (see supplementary information),
\begin{eqnarray}
& & \frac{dn}{dt} = - A_{f}n_{f}n^{2}(1-F_{f}) - A_{s}n_{s}n^{2}(1-F_{s}) + gI(t) \label{eq:2} \\
& & \frac{dp}{dt} = - B_{f}n_{f} n p F_{f} - B_{s}n_{s} n p F_{s} + gI(t) \label{eq:3} \\
& & n_{f/s}\frac{dF_{f/s}}{dt} = A_{f/s}n_{f/s}n^{2}(1-F_{f/s}) -   B_{f/s}n_{f/s} n p F_{f/s} \label{eq:4} 
\end{eqnarray} 
Here, $n$ and $p$ are the electron and hole densities and include both free electrons and holes as well as bound electrons and holes (excitons) (see supplementary information), $n_{f}(n_{s})$ is the density of fast (slow) defect levels, and $F_{f} (F_{s})$ is the electron occupation of the fast (slow) defect level. $A_{f/s}$ ($B_{f/s}$) is the rate constant for the Auger capture of electrons (holes) by the defects. $I(t)$ is the pump pulse intensity (units: $\mu$J/cm$^{2}$-s) and $g$ equals $\sim$2.5$\times$10$^{11}$ 1/$\mu$J (from measurements). Change in the probe pulse transmission through the photoexcited sample is assumed to be given by (\ref{eq:1}) and (\ref{eq:drude}). The essential dynamics captured by the above Equations are presented in Figure \ref{fig:cartoon} and consist of three main steps corresponding to the three temporal regions in Figure \ref{fig:power}: (I) After photoexcitation, the carriers thermalize and cool. This is assumed to happen instantly in our model. (II) Most of the holes, followed by the electrons, are captured by the fast defects within a few picoseconds. This fast capture process is responsible for the fast time scale observed in region (II) in Figure \ref{fig:power}. A small fraction of the photogenerated holes is also captured by the slow defects. (III) After all the photoexcited holes have been captured and the electrons have filled the fast traps completely, the remaining photoexcited electrons are captured by the slow defects. This last step is slow and is responsible for the slow time scale observed in region (III) in Figure \ref{fig:power}. The choice of the values of the parameters in Equations (\ref{eq:2})-(\ref{eq:4}) can be aided by the data. Assuming small pump fluence and knowing the equilibrium electron density $n_{o}$ and the fact that immediately after photoexcitation $F_{f} \approx F_{s}\approx 1$, the value of the product $B_{f}n_{f}$ in (\ref{eq:3}) is chosen to match the time scale of the fast transient in region (II) in Figure \ref{fig:power}. The value of the product $B_{s}n_{s}$ is chosen to adjust the fraction of the holes that is captured by the slow defects in region (II) in order to fit the relative value of $|\Delta T/T|$ in the beginning of region (III) compared to the peak  value. From our data, $B_{s}n_{s}$ should be 7.5-8 times smaller than $B_{f}n_{f}$. Knowing the hole density captured by the slow defects in region (II) (which equals $n_{s}(1-F_{s})$ at the end of region (II)), the value of $A_{s}$ is chosen to match the time scale of the slow transient in region (III) in Figure \ref{fig:power}. Finally, the value of $A_{f}$ is chosen to match the dependence of the peak value of $|\Delta T/T|$ on the pump fluence, as show in Figure \ref{fig:power_scaling}(b). Parameter values obtained this way are shown in Table 1. Once the parameter values have been chosen in this way, we find that the simulations fit the data very well for a small range of values of the defect densities, $n_{f}$ and $n_{s}$, as indicated in Table 1. These defect density values compare well with the theoretically predicted and observed point defect densities in MoS$_{2}$~\cite{Sofo04,Komsa12,Zhou13,Kim14,Guinea14,Robertson13,Hao13,VanDerZande13}. In the supplementary information, we show that the theoretically estimated values of the Auger capture rate constants are in the same range as the values given in Table 1. 
\begin{table} 
\begin{center}
\begin{tabular}{|l|r|}
\hline
$B_{f}n_{f}$& $0.73\pm0.05$ cm$^{2}$/s\\ \hline
$B_{s}n_{s}$& $9.3\pm1\times10^{-2}$  cm$^{2}$/s \\ \hline
$A_{s}$& $9.5\pm1\times10^{-15}$  cm$^{4}$/s \\ \hline
$A_{f}$& $(1.0\pm 0.2)B_{f}$ \\ \hline
$n_{f}$& $0.3\times10^{12}$ to $10^{12}$ 1/cm$^{2}$ \\ \hline
$n_{s}$& $10^{12}$ to $2.0\times10^{13}$ 1/cm$^{2}$ \\ \hline
$n_{o}$ & $2.0\times10^{12}$ 1/cm$^{2}$ \\ \hline
\end{tabular}
\caption{Parameter values used in the simulations to fit the transient data.}
\end{center}
\end{table}
\begin{figure}[tbh]
  \centering
  \includegraphics[width=.6\textwidth]{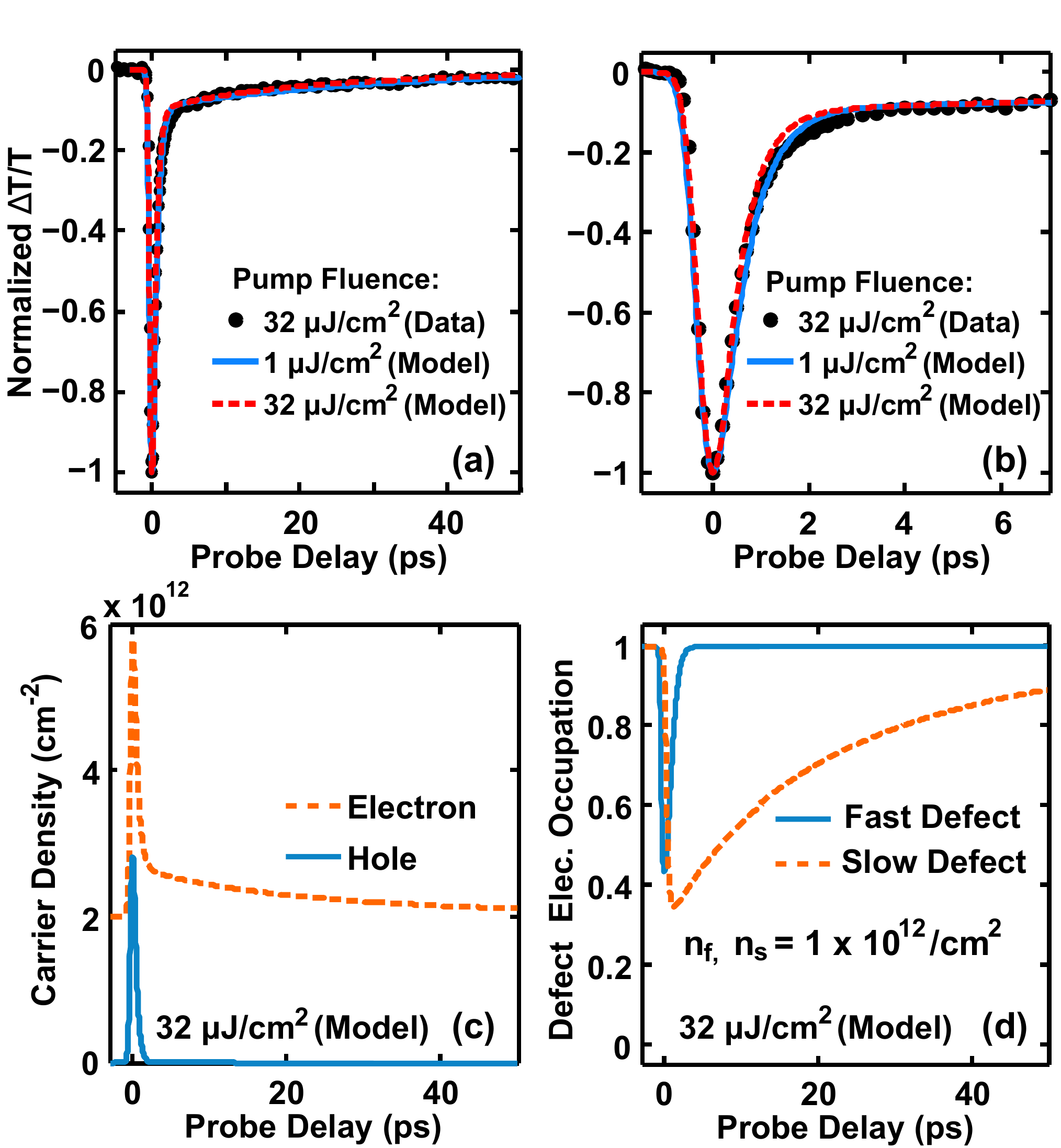}
  \caption
  {(a, b) The measured and calculated $\Delta T/T$ (pump fluence normalized) are plotted as a function of the probe delay. The plot in (a) is the same as the plot in (b) and shows the details at small times. The calculations are done for two different and extreme values of the pump fluence, 1 and 32 $\mu$J/cm$^{2}$, used in our experiments. The Auger carrier capture model reproduces all the time scales observed in the measurements over the entire range of the pump fluence values used. (c) The simulated electron and hole densities are plotted as a function of time after photoexcitation for 32 $\mu$J/cm$^{2}$ pump fluence. (d) The simulated electron occupations, $F_{f}$ and $F_{s}$, of the fast and slow traps are plotted as a function of time after photoexcitation for 32 $\mu$J/cm$^{2}$ pump fluence.}
   \label{fig:sim}
\end{figure}

The agreement between the simulation results for $\Delta T/T$ and the measurements are presented in Figures \ref{fig:sim} and \ref{fig:power_scaling}(b). In Figures \ref{fig:sim}(a,b), the measured and calculated $\Delta T/T$ (pump fluence normalized) are plotted as a function of the probe delay. The calculations are done for two different and extreme values of the pump fluence, 1 and 32 $\mu$J/cm$^{2}$, used in our experiments. The model not only reproduces the very different time scales observed in $\Delta T/T$ measurements in regions (II) and (III), it achieves a very good agreement with the data over the entire range of the pump fluence values, 1 to 32 $\mu$J/cm$^{2}$, used in our experiments. The near exponential appearance of the measured transient in region (III) (see Figure \ref{fig:log_data}) is also reproduced by our model despite the fact that the Auger capture rates are not linear functions of the carrier density. This can be understood as follows. In region (III), if at any time the hole density in the slow traps equals $n_{s}(1-F_{s})$ then this is also equal to the photoexcited electron density $n-n_{o}$ left in the conduction band. Therefore, in region (III) the rate equation for the electron density becomes,   
\begin{equation}
\frac{dn}{dt} \approx - A_{s}n_{s}n^{2}(1-F_{s}) = - A_{s}n^{2}(n-n_{o}) \approx - A_{s}n_{o}^{2}(n-n_{o}) \label{eq:2b} 
\end{equation} 
The last approximate equality above follows from the fact that in region (III) the remaining photoexcited electron density $n-n_{o}$ is much smaller than the doping density $n_{o}$ for all pump fluence values used in our experiments. The above Equation shows that in region (III) the transient will behave almost like a decaying exponential with an inverse time constant given by $A_{s}n_{o}^{2}$. Figures \ref{fig:sim}(b) and (c) show the calculated carrier densities and the defect occupations for the maximum pump fluence of 32 $\mu$J/cm$^{2}$ used in our experiments. The carrier and the defect state dynamics depicted in Figure \ref{fig:cartoon} are reproduced by the model using the parameter values given in Table 1. From the computed carrier densities, one can obtain the scaling of the peak value of $|\Delta T/T|$ with the pump fluence. Figure \ref{fig:power_scaling}(b) shows the measured and the calculated dependence of the peak value of $|\Delta T/T|$ on the pump fluence, and the model is again seen to agree well with the measurements. Finally, the calculated values of $\Delta T/T$, using (\ref{eq:drude}), agree very well with the measurements if one assumes a carrier mobility, $e\tau/(0.5m_{o})$, of $\sim$35 cm$^{2}$/V-s, which is in the range of the values reported for exfoliated MoS$_{2}$ monolayers~\cite{Jariwala13}.

\noindent
\textbf{Conclusion:} In this work, we presented experimental results on the ultrafast dynamics of photoexcited carriers in monolayers of MoS$_{2}$ and showed that defect assisted electron-hole recombination, in which carrier capture by defects occurs via Auger scattering, explains our observations very well. Based on the dependence of the measured data on the temperature and the pump fluence, we ruled out other mechanisms of non-radiative recombination and carrier capture by defects. Strong Coulomb interactions in two dimensional materials make Auger scattering effective. Our results will be helpful in understanding and evaluating the performance of MoS$_2$-based electronic and optoelectronic devices. Finally, we note here that our measurements might not have detected charge trapping dynamics occurring on much longer time scales ($\gg$10 ns) recently observed in MoS$_{2}$ photoconductive devices\cite{Cho14}.

\noindent
\textbf{Acknowledgments:} The authors would like to acknowledge helpful discussions with Jared H. Strait, Michael G. Spencer, and Paul L. McEuen, and support from CCMR under NSF grant number DMR-1120296, AFOSR-MURI under grant number FA9550-09-1-0705, ONR under grant number N00014-12-1-0072, and the Cornell Center for Nanoscale Systems funded by NSF.

\newpage

\noindent
\huge{Supplementary Information}
\normalsize
\newline

\begin{figure}[tbh]
  \centering
  \includegraphics[width=.6\textwidth]{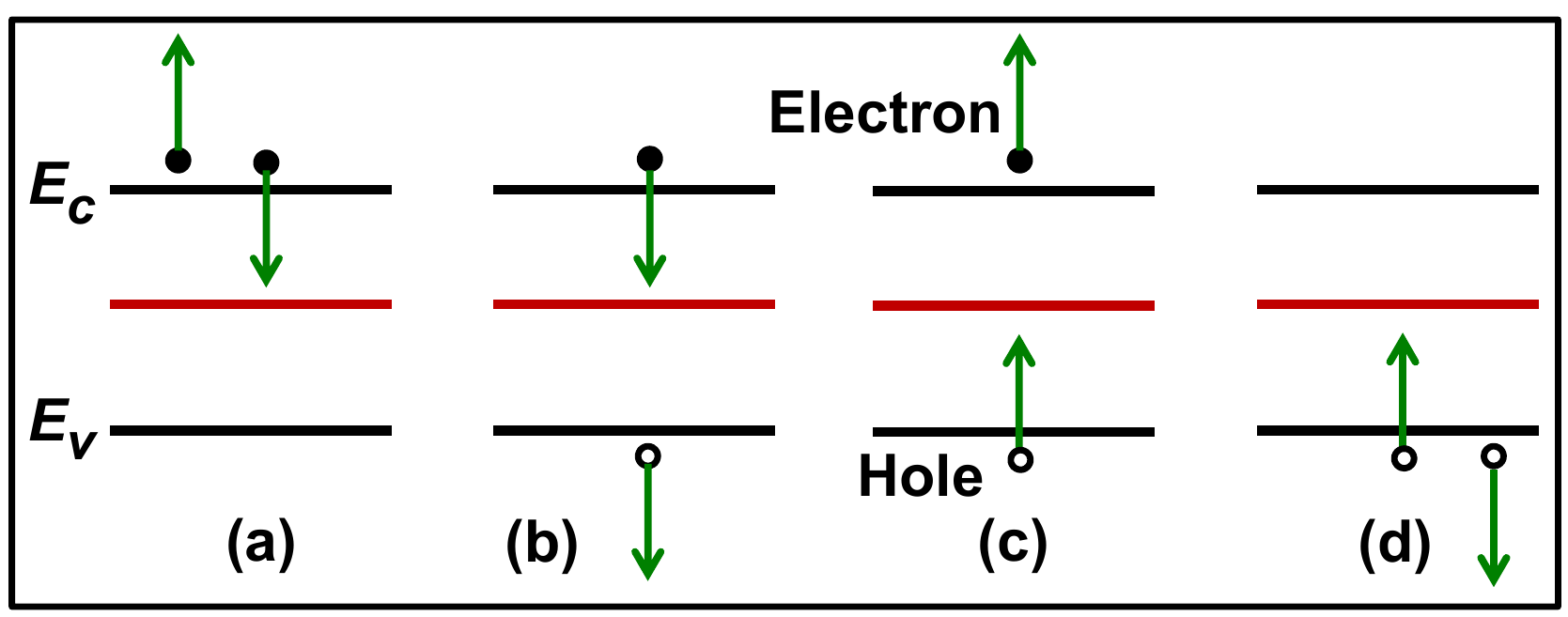}
  \caption{Four basic Auger processes for the capture of electrons ((a) and (b)) and holes ((c) and (d)) at defects are depicted~\cite{Jaros80,Landsberg92,Landsberg80}.}
  \label{fig:auger_supp}
\end{figure}

\noindent
\textbf{Auger Carrier Capture Rates in an Electron-Hole Plasma in MoS$_{2}$ Monolayer:} Four basic Auger processes for the capture of electrons ((a) and (b)) and holes ((c) and (d)) at defects are depicted in Figure \ref{fig:auger_supp}. In this Section, we derive expressions for the Auger carrier capture rates in two dimensional materials like MoS$_{2}$. We make order of magnitude estimates for the rates and show that strong Coulomb interactions in two dimensional materials can make Auger capture rates fairly large and consistent in magnitude with our experimental observations. For simplicity, we will look at process (c) in detail in which a hole scatters off an electron and is captured by a deep defect and the electron is scattered to a higher energy. The rates for all other processes can be calculated in a similar manner. We assume a defect level at an energy $E_{d}$ above the valence band maxima at the $K$ and $K'$ points in the first Brillouin zone of MoS$_{2}$. The wavefunction of the electron in the defect state can be written most generally as,
\begin{equation}
\phi_{d}(\vec{r}) = \sum_{n,\vec{k}} c_{n}(\vec{k}) \psi_{n,\vec{k}}(\vec{r}) =   \sum_{n} A_{n}(\vec{r}) u_{n,\vec{k_{o}}}(\vec{r})
\end{equation}
Here, the sum over $n$ runs over all the energy bands, $\psi_{n,\vec{k}}(\vec{r})$ are the Bloch functions, $u_{n,\vec{k}_{o}}(\vec{r})$ are the periodic part of the Bloch functions evaluated for $\vec{k} = \vec{k}_{o}$, where $\vec{k}_{o}$ could be the location of either one of the $K$ or $K'$ points under consideration. The second equality follows in the spirit of multi-band effective mass theory. A deep level, unlike a shallow level, is likely to have contributions from multiple bands~\cite{Landsberg92,Landsberg80}. The functions $ A_{n}(\vec{r})$ are assumed to be localized near the defect. The relevant term in the Coulomb interaction that describes the process in Figure \ref{fig:auger_supp}(b) can be written as,
\begin{equation}
H =  \frac{1}{A^{3/2}} \sum_{\vec{q},\vec{k},\vec{k}',j} V(\vec{q}) F(\vec{k},\vec{k}',\vec{q}) c^{\dagger}_{\vec{k}+\vec{q}} b^{\dagger}_{\vec{k}'} d_{j}c_{\vec{k}} 
\end{equation}
We have suppressed spin and valley indices in the above expression for simplicity. $b_{\vec{k}}$, $c_{\vec{k}}$, and $d_{j}$  are the destruction operators for states in the valence band, conduction band, and the defect. $j$ runs over all the defects in the crystal. $V(q)$, given by $e^{2}/(2\epsilon_{o} \epsilon(\vec{q})q)$, is the 2D Coulomb potential and the dielectric constant $\epsilon(\vec{q})$ appropriate for MoS$_{2}$ is given by Zhang et~al.~\cite{Changjian14}. $A$ is the area of the sample. The overlap factor $F(\vec{k},\vec{k}',\vec{q})$ equals,
\begin{equation}
F(\vec{k},\vec{k}',\vec{q}) = \frac{1}{A} \int d^{2}\vec{r} \, u^{*}_{c,\vec{k}+\vec{q}}(\vec{r}) u_{c,\vec{k}}(\vec{r}) \,\, \sum_{n} \frac{1}{A} \int d^{2}\vec{r} \, u^{*}_{v,\vec{k}'}(\vec{r}) u_{n,\vec{k}_{o}}(\vec{r}) \,\, \int d^{2}\vec{r} \, A_{n}(\vec{r}) e^{-i(\vec{k}'+\vec{q}).\vec{r}} \label{eq:R0}
\end{equation}
The initial many body state $|i\rangle$ is the electron-hole plasma created after photoexcitation and consists of free as well as bound (excitons) carriers. An expression for the defect capture rate can be derived that incorporates correlations between electrons and holes. Using Fermi's golden rule, the hole capture rate $R$ (units: 1/cm$^{2}$-s) for the process depicted in Figure \ref{fig:auger_supp}(b) can be written as,
\begin{equation}
R = \frac{2\pi}{A\hbar}\sum_{f} \left| \langle f | H |i\rangle \right|^{2}\delta(E_{f}-E_{i}) \label{eq:R1} 
\end{equation}
The squared matrix element in the above expression is proportional to the many body correlation function,
\begin{equation}
\frac{1}{A^{2}}\sum_{\vec{k}_{1},\vec{k}'_{1},\vec{k}_{2},\vec{k}'_{2}} \langle i|   c^{\dagger}_{\vec{k}_{1}} b_{\vec{k}'_{1}} b^{\dagger}_{\vec{k}'_{2}} c_{\vec{k}_{2}} |i \rangle
\end{equation}
As shown by Kira et~al.~\cite{Kira06}, the above correlation function can be generally approximated as, $np(1 + g(\vec{r}=0))$, where $n$ and $p$ are the total electron and hole densities, including electrons and holes present in the form of bound excitons, and the dimensionless function $g(\vec{r}=0)$ expresses the enhancement in the probability of finding an electron and a hole at the same location in space due to correlations~\cite{Kira06}. If the electrons and holes are completely uncorrelated then $g(\vec{r}=0) \approx 0$. On the other hand, if all the electrons and holes are bound and are in the lowest exciton level (1s) with relative coordinate wavefunction $\phi^{R}_{\rm 1s}(\vec{r})$, then $g(\vec{r}=0) \propto  |\phi^{R}_{\rm 1s}(\vec{r}=0)|^{2}$~\cite{Kira06}. Given that the exciton radii in monolayer MoS$_{2}$ is 7-9 A$^\circ$ (Zhang et~al.\cite{Changjian14}), $g(\vec{r}=0)$ can be larger than $\sim$100. Electron-hole correlations induced by Coulomb interactions can therefore drastically increase the Auger carrier capture rates.

In order to compute the capture rate in (\ref{eq:R1}), we need to specify the initial and final energies, $E_{i}$ and $E_{f}$, respectively. With reference to the wavevectors in the correlation function $\langle i|   c^{\dagger}_{\vec{k}_{1}} b_{\vec{k}'_{1}} b^{\dagger}_{\vec{k}'_{2}} c_{\vec{k}_{2}} |i \rangle$, and assuming a momentum $\vec{q}$ is transferred to an electron in the conduction band in the process depicted in Figure \ref{fig:auger_supp}(b), we obtain,
\begin{equation}
E_{f} - E_{i} =  \frac{\hbar^{2}|\vec{k}_{2}+\vec{q}|^2}{2m_{e}} + E_{\alpha} - \frac{\hbar^{2}|\vec{k}_{2}-\vec{k}'_{2}|^2}{2(m_{e}+m_{h})} - E_{d} \label{eq:R2}
\end{equation} 
If the initial electron and hole pair were bound, $E_{\alpha}$ would be the binding energy associated with the exciton wavefunction $\phi_{\alpha}^{R}(\vec{r})$. If the pair were not bound, $E_{\alpha}$ can be approximated as the relative motion kinetic energy of the pair,
\begin{equation}
E_{\alpha} \approx - \frac{\hbar^{2}|\vec{k}_{2}m_{r}/m_{e} + \vec{k}'_{2}m_{r}/m_{h}|^2}{2m_{r}}
\end{equation}
The term $\hbar^{2}|\vec{k}_{2}-\vec{k}'_{2}|^2/2(m_{e}+m_{h})$ is the center of mass energy of the initial electron and hole pair, and $\hbar^{2}|\vec{k}_{2}+\vec{q}|^2/2m_{e}$ is the energy of the scattered electron. Assuming that the defect level is deep, and $E_{d}$ is very large, $\vec{q}$ will be much larger than $\vec{k}_{2}$ or $\vec{k}'_{2}$. Therefore, we approximate the delta function appearing in (\ref{eq:R1}) as,
\begin{equation}
\delta(E_{f} - E_{i}) \approx \delta(\hbar^{2}q^{2}/2m_{e}-E_{d})
\end{equation}
We also approximate the overlap factor $F(\vec{k}_{2},\vec{k}'_{2},\vec{q})$ as, 
\begin{equation}
|F(\vec{k}_{2},\vec{k}'_{2},\vec{q})|^{2}  \approx  \left| \int d^{2}\vec{r} \, A_{v}(\vec{r}) e^{-i(\vec{q}).\vec{r}} \right|^{2} = |A_{v}(\vec{q})|^{2}  
\end{equation}
Assuming a defect density $n_{d}$, and an average defect occupation of $F_{d}$, and including spin and valley degeneracies, the hole capture rate in (\ref{eq:R1}) can be written in a simple form,
\begin{equation}
R \approx B n_{d} n p F_{d} =   \frac{m_{e}}{\hbar^{3}} |V(q_{d})|^{2}(1 + g(\vec{r}=0))|A_{v}(\vec{q_{d}})|^{2} n_{d} np F_{d} \label{eq:Rf}
\end{equation}
where $q_{d} = \sqrt{2m_{e}(E_{d})}/\hbar$. If the defect wavefunction is assumed to have the spherically symmetric shape of an s-orbital with radius $a$ then,
\begin{equation}
A_{v}(\vec{r}) = \sqrt{\frac{2}{\pi a^{2}}} e^{-r/a} \;\;\;\;\; |A_{v}(\vec{q_{d}})|^{2}  = \frac{8\pi a^{2}}{[1 + (q_{d}a)^2]^{3}}
\end{equation}
In order to make an order of magnitude estimate of the value of the rate constant $B$ in (\ref{eq:Rf}), we assume a defect wavefunction radius of $\sim$4 A$^\circ$, which is appropriate for highly localized deep defects~\cite{Landsberg92}, and a defect energy $\sim$1 eV above the valence band edge ($E_{d} = 1$ eV). Assuming $m_{h} = m_{e} = 0.5m_{o}$~\cite{Changjian14}, and using the expression for the wavevector dependent dielectric constant for a MoS$_{2}$ monolayer on a quartz substrate given by Zhang et~al.\cite{Changjian14}, we obtain values of $B$ equal to $2.2 \times 10^{-13}$ cm$^{4}$/s and $2.2 \times 10^{-11}$ cm$^{4}$/s for assumed values of $g(\vec{r}=0)$ equal to 0 and 100, respectively. In comparison, the largest experimentally determined values in this work (see value of $B_{f}$ for the fast traps in Table 1 in the text) are in the range $7.3\times10^{-13}$- $2.4\times 10^{-12}$  cm$^{4}$/s (taking into account the uncertainty in the sample defect density listed in Table 1). Therefore, the rate constants for Auger carrier capture processes in MoS$_{2}$ determined from the theory agree well with the experimental data. Interestingly, the smallest and the largest values of $g(\vec{r}=0)$, used as a fitting parameter, needed to obtain an exact match between the theory and our experimental data are $\sim$2.3 and $\sim$10, respectively.

Similar calculations shows that the rate of the other capture processes in Figure \ref{fig:auger_supp} are,
\begin{eqnarray}
& (a)&  \,\,\, R \approx A n_{d} n^2 (1-F_{d}) =   \frac{3}{4} \frac{m_{e}}{\hbar^{3}} |V(q_{d})|^{2} |A_{c}(\vec{q_{d}})|^{2} n_{d} n^{2} (1-F_{d}) \label{eq:Ra} \\
& (b) &  \,\,\, R \approx A n_{d} n p (1-F_{d}) =   \frac{m_{h}}{\hbar^{3}} |V(q_{d})|^{2}(1 + g(\vec{r}=0))|A_{c}(\vec{q_{d}})|^{2} n_{d} np (1-F_{d}) \label{eq:Rb} \\
& (c) &  \,\,\, R \approx B n_{d} n p F_{d} =   \frac{m_{e}}{\hbar^{3}} |V(q_{d})|^{2}(1 + g(\vec{r}=0))|A_{v}(\vec{q_{d}})|^{2} n_{d} np F_{d} \label{eq:Rc} \\
& (d) &  \,\,\, R \approx B n_{d} p^2 F_{d} =   \frac{3}{4} \frac{m_{h}}{\hbar^{3}} |V(q_{d})|^{2} |A_{v}(\vec{q_{d}})|^{2} n_{d} p^{2} F_{d}  \label{eq:Rd}
\end{eqnarray}
In (a) and (b) the defect energy $E_{d}$ is measured from the conduction band bottom, and in (c) and (d) the defect energy $E_{d}$ is measured from the valence band top.
\newline

\noindent
\textbf{Intraband Absorption by Free Carriers and Excitons:} In this Section, we derive an expression for the excitonic contribution to the intraband absorption and show that at optical frequencies much higher than the exciton binding energies and much lower than the optical bandgap, the intraband conductivity of excitons looks similar to the intraband conductivity of free carriers. We assume a MoS$_{2}$ monolayer with free electron density $n_{f}$, free hole density $p_{f}$, and bound exciton density $n_{ex}$. The total electron density $n$ is $n_{f} + n_{ex}$ and the total hole density $p$ is $p_{f} + n_{ex}$. The total intraband conductivity $\sigma(\omega)$ can be written as $\sigma_{f}(\omega) + \sigma_{b}(\omega)$, where $\sigma_{f}(\omega)$ is the contribution from the free carriers and $\sigma_{b}(\omega)$ is the contribution from the bound excitons. We now show that the contribution to the real part of the intraband conductivity from all the carriers, free and bound, at the frequencies of interest can be written as,
\begin{equation}
\sigma_{r}(\omega) \approx \left( \frac{n}{m_{e}} + \frac{p}{m_{h}} \right) \frac{e^{2}\tau}{1 + \omega^{2} \tau^{2}} \label{eq:drude2}   
\end{equation}
The real part of the intraband conductivity contribution from the free carriers must satisfy the partial sum rule~\cite{qtll05},
\begin{equation}
\int_{0}^{\infty} \sigma_{rf}(\omega) d\omega = \frac{\pi e^{2}}{2} \left(  \frac{n_{f}}{m_{e}} + \frac{p_{f}}{m_{h}} \right)
\end{equation}
In addition, the high frequency limit of the imaginary part of the conductivity $\sigma_{if}(\omega)$ follows from the Lehmann representation of the current correlation function~\cite{qtll05},
\begin{equation}
\lim_{\omega \to \infty} \sigma_{if}(\omega) = i \left( \frac{n_{f}}{m_{e}} + \frac{p_{f}}{m_{h}} \right) \frac{e^{2}}{\omega}
\end{equation}
An expression that satisfies both the above conditions is given by the Drude form,
\begin{equation}
\sigma_{f}(\omega) = i\left( \frac{n_{f}}{m_{e}} + \frac{p_{f}}{m_{h}} \right) \frac{e^{2}}{\omega + i/\tau} \label{eq:drude3}
\end{equation}
The above well known expression for the Drude form of the free carrier conductivity can be derived in many different ways~\cite{qtll05}. To find the intraband conductivity of the excitons we start from the relative coordinate operator for the exciton,
\begin{equation}
\hat{\vec{r}} =  \hat{\vec{r}}_{h} -  \hat{\vec{r}}_{e}
\end{equation}
where $\hat{\vec{r}}_{h}$ and  $\hat{\vec{r}}_{e}$ are electron and hole position operators. We have, 
\begin{equation}
\left[\hat{\vec{r}}.\hat{n},\hat{H}\right] = i\hbar\frac{\hat{\vec{p}}.\hat{n}}{m_{r}}
\end{equation}
where $\hat{\vec{p}}$ is the relative momentum conjugate to $\hat{\vec{r}}$, $\hat{n}$ is any unit vector, and $m_{r}$ is the reduced electron-hole mass,
\begin{equation}
\frac{1}{m_{r}} =  \frac{1}{m_{e}} + \frac{1}{m_{h}}
\end{equation}
It follows that,
\begin{equation}
\left[\hat{\vec{r}}.\hat{n},\left[\hat{\vec{r}}.\hat{n},\hat{H}\right]\right] = -\frac{\hbar^{2}}{m_{r}}
\end{equation}
Suppose $|\alpha \rangle$ represent all exciton states, tightly bound as well as ionized, with energies $E_{\alpha}$. Taking the matrix elements of the above commutator equation with the exciton states we get,
\begin{equation}
\sum_{\beta} (E_{\beta} - E_{\alpha}) |\langle \alpha | \hat{\vec{r}}.\hat{n} | \beta \rangle|^{2} = \frac{\hbar^{2}}{2m_{r}} \label{eq:TRK}
\end{equation} 
The above expression is a modification of the well known Thomas-Reich-Kuhn oscillator strength sum rule. Most of the oscillator strength on the left hand side comes from the terms in which both $|\alpha \rangle$ and $|\beta \rangle$ are bound exciton states or low energy ionized states. Now consider an exciton gas with density $n_{ex}$ in which the state $|\alpha \rangle$ is occupied with probability $\rho_{\alpha}$. We only consider bound exciton states to be occupied here since the contribution from the occupied exciton states that are ionized has been included in the intraband contribution from the free carriers considered above. The interaction between an exciton and a classical electromagnetic field of frequency $\omega$ is given by the dipole operator $-e\hat{\vec{r}}.\hat{n} E(t)$, where $\hat{n}$ is the field polarization unit vector. The optical intraband conductivity of the system can be found easily using standard linear response techniques and comes out to be,
\begin{equation}
\sigma_{b}(\omega) =   ie^{2} n_{ex} \sum_{\alpha, \beta} |\langle \beta | \hat{\vec{r}}.\hat{n} | \alpha \rangle|^{2} \rho_{\alpha} \left[ \frac{2\omega(E_{\beta} -E_{\alpha})}{(\hbar \omega + i\hbar/\tau)^{2} - (E_{\beta} -E_{\alpha})^{2}} \right] \label{eq:excond}
\end{equation}
We have introduced a phenomenological damping parameter $\tau$. Using the sum rule given in (\ref{eq:TRK}), the bound exciton conductivity $\sigma_{b}(\omega)$ is found to satisfy the conductivity sum rule,
\begin{equation}
\int_{0}^{\infty} \sigma_{rb}(\omega) d\omega = \frac{\pi e^{2}}{2} \left(  \frac{n_{ex}}{m_{r}} \right)
\end{equation}
Since most of the oscillator strength in the sum rule comes from the bound states or the low energy ionized states, one can take the large frequency limit of the expression in (\ref{eq:excond}) and obtain,
\begin{equation}
\sigma_{b}(\omega) \approx   i \frac{e^{2}}{\hbar^{2}} n_{ex} \sum_{\alpha, \beta} |\langle \beta | \hat{\vec{r}}.\hat{n} | \alpha \rangle|^{2} \rho_{\alpha} \left[ \frac{2 (E_{\beta} -E_{\alpha})}{\omega + i/\tau} \right] \label{eq:excond2}
\end{equation}
The above expression is valid for frequencies $\omega$ much higher than the exciton binding energies but much lower than the material optical bandgap. Using the sum rule given in (\ref{eq:TRK}), we obtain,
\begin{equation}
 \sigma_{b}(\omega) \approx   i \left( \frac{n_{ex}}{m_{r}} \right) \frac{e^{2}}{\omega + i/\tau}  =  i \left( \frac{n_{ex}}{m_{e}} + \frac{n_{ex}}{m_{h}} \right) \frac{e^{2}}{\omega + i/\tau} \label{eq:excond3}
\end{equation}
Under the assumption that the damping parameters $\tau$, appearing in the conductivity expressions for free carriers and bound excitons, are approximately the same, we can add $\sigma_{b}(\omega)$ from (\ref{eq:excond3}) and $\sigma_{f}(\omega)$ from (\ref{eq:drude3}), and obtain the simple expression for the intraband conductivity of the sample at the frequencies of interest,
\begin{equation}
\sigma(\omega) = \sigma_{f}(\omega) + \sigma_{b}(\omega) \approx  i \left( \frac{n}{m_{e}} + \frac{p}{m_{h}} \right) \frac{e^{2}}{\omega + i/\tau}
\end{equation}
Note that $n$ and $p$ above are now the total electron and hole densities including free carriers and bound carriers (excitons). The desired expression in (\ref{eq:drude2}) follows immediately by taking the real part of the above expression.   
\newline

\begin{figure}[tbh]
  \centering
  \includegraphics[width=.8\textwidth]{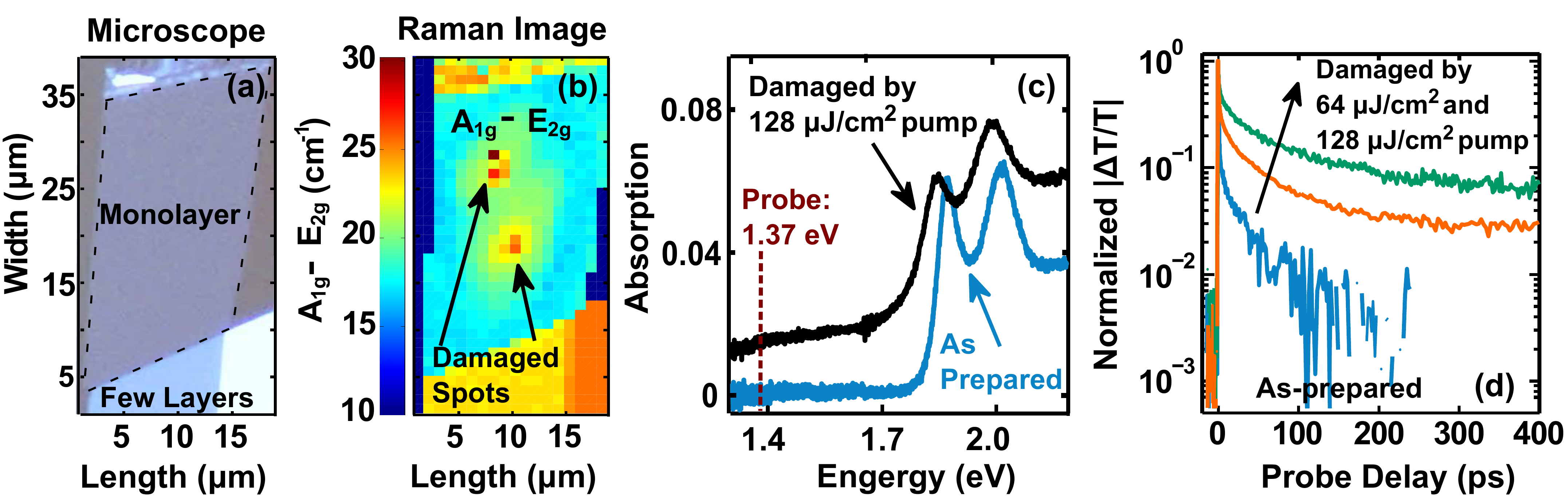}
  \caption
  {(a) Optical microscope image of a sample damaged by a  $\sim$128 $\mu$J/cm$^{2}$ pump pulse is shown. No sign of damage is visible. (b) A scanned Raman image corresponding to the energy splitting between the $A_{1g}$ and $E_{2g}$ modes is shown for a sample damaged by a $\sim$128 $\mu$J/cm$^{2}$ pump pulse. (c) A sample damaged by a  $\sim$128 $\mu$J/cm$^{2}$ pump pulse shows large absorption throughout the bandgap. (d) The transmission of the probe pulse at 905 nm wavelength in a damaged sample gets overwhelmed by the increased absorption inside the bandgap and, consequently, the dynamics observed by the probe pulse become completely different compared to the dynamics observed in an undamaged sample. A damaged sample shows a large slow component in the transient that has a time scale much longer than any of the time scales observed in an undamaged sample. The magnitude of the slow component in the measured transient is larger in samples damaged with a higher pump fluence. }
     \label{fig:damaged}
\end{figure}

\noindent
\textbf{Optically Induced Sample Damage and its Effect on the Optical Properties and on the Measured Dynamics:} In our experiments we found that exfoliated MoS$_{2}$ monolayers (SPI Supplies and 2D Semiconductors) could easily get damaged permanently when pump fluences in excess of $\sim$50 $\mu$J/cm$^{2}$ were used (452 nm wavelength). Once damaged in this way, the optical characteristics of the sample would change completely. So care needed to be exercised in ensuring that samples were not damaged during pump-probe, photoluminescence, or absorption measurements. An optical microscope image and a Raman spectroscopy image of a sample damaged by pump pulses is shown in Figures \ref{fig:damaged}(a) and (b). The energy splitting between $A_{1g}$ and $E_{2g}$ mode is much larger at the damaged spots compared to the surrounding intact areas. Similar enhancement of the mode splitting in optically damaged samples has been reported previously~\cite{Steele12}. In addition, damaged samples exhibited large optical absorption throughout the bandgap, as shown in Figure \ref{fig:damaged}(c), indicating creating of midgap states. Finally, the transmission of the probe pulse at 905 nm wavelength in a damaged sample gets overwhelmed by the increased absorption inside the bandgap and, consequently, the dynamics observed by the probe pulse become completely different compared to the dynamics observed in an undamaged sample. A damaged sample shows a large slow component in the transient that has a time scale much longer than any of the time scales observed in an undamaged sample, as shown in Figure \ref{fig:damaged}(d). The magnitude of the slow component in the measured transient is larger in samples damaged with a higher pump fluence. We strongly feel that it is very important that the absence or presence of optically induced sample damage is checked before/after optical measurements in order to ensure that reliable data has been obtained.

\bibliography{m_opop_2}

\end{document}